\documentclass{PoS}

\title{Hadron spectra from overlap fermions on HISQ gauge configurations.}

\ShortTitle{Hadron spectra from overlap fermions on HISQ gauge configurations.}

\author{S. Basak$^a$, S. Datta$^b$, A. T. Lytle$^b$, Padmanath M.$^b$, P. Majumdar$^c$, and 
{\speaker{N. Mathur}}$^b$}

\author{{\hspace*{1.2in}}(Indian Lattice Gauge Theory Initiative)\\\\
\llap{$^a$}School of Physical Sciences, National Institute of Science Education 
and Research, Bhubaneswar 751 005, India\\
\llap{$^b$}Department of Theoretical Physics, Tata Institute of Fundamental Research, Homi Bhabha Road, Mumbai 400005, India\\
\llap{$^c$}Department of Theoretical Physics, Indian Association for the Cultivation of Science, Kolkata 700032, India.
\\
        E-mail: \email{nilmani@theory.tifr.res.in}}

\abstract{Adopting a mixed action approach, we report here results on hadron spectra containing one or more charm quarks. 
We use overlap valence quarks on a background of 2+1+1 flavor HISQ gauge configurations generated by the MILC collaboration.
 We also study the ratio of leptonic decay constants, $f_{D_s^*}/f_{D_s}$. Results are obtained at two lattice spacings.}

\FullConference{31st International Symposium on Lattice Field Theory LATTICE 2013\\July 29-August 3, 2013\\Mainz, Germany}

\begin{document}
\vspace{-0.2in}
\section{Introduction}
\vspace{-0.05in}
Recently there has been a resurgence of interest in heavy hadron
spectroscopy with the discovery of numerous hadrons with one or more
heavy quarks. Results from the LHC and future charm-bottom factories are
expected to add to the excitement in this field in the near future.
However, the study of heavy hadrons using lattice QCD has an
inherent problem since at these masses, with currently available
lattices, the condition $am << 1$ in general is not satisfied which
leads to larger systematic errors. Though NRQCD is successful in
studying bottom quark it is not so clear whether one can use that to
study hadrons with one or more charm quarks. Relativistic heavy quark
actions, where all $O((am)^n)$ corrections are systematically removed,
are becoming increasingly
popular~\cite{ElKhadra:1996mp,hvquark_action}.

In this work we have adopted a mixed action approach by the using overlap
action~\cite{overlap} for valence quarks on a background of 2+1+1 flavours HISQ gauge
configurations~\cite{HISQ_MILC}. The overlap action is automatically ${\cal{O}}(ma)$
improved; it also offers various simplifications in studies of decay
constants.  An aim of this study is to investigate the behavior of the
overlap action in the regime $ma \lesssim 1$. The overlap action also
has some desirable features computationally, such as the adaptation of
multi mass algorithms~\cite{overlap_multimass}.  However, using
overlap action for the dynamical quarks is still prohibitively costly,
except with fixed topology~\cite{overlap_dynamical}. Therefore, for
the gauge configurations we have used the large set of 2+1+1 flavours
configurations generated by the MILC lattice collaboration
\cite{HISQ_MILC} with the one-loop, tadpole improved Symanzik gauge
action and the highly improved staggered quark (HISQ) fermion
action~\cite{HISQ_action}. Taste violations in the HISQ action were
found to be small~\cite{HISQ_action}. A similar mixed action approach has been taken by the $\chi QCD$
collaboration using overlap valence quarks on 2+1 flavours
dynamical domain wall gauge configurations~\cite{overlap_chiQCD1}. 

In this report, we present our preliminary results on charm and
strange hadron spectra as well as leptonic decay constants for $D_s$
and $D^*_s$ mesons, using the above-mentioned mixed action
approach. This is an update of our ongoing study; earlier results were
reported in Ref.~\cite{Basak:2012py}.

\vspace{-0.1in}
\section{Numerical details}
\vspace{-0.05in}
We used two sets of dynamical 2+1+1 flavours HISQ lattice ensembles,
generated by the MILC collaboration : a set of $32^3 \times 96$
lattices at gauge coupling $10/g^2 = 6.30$ and another
set of $48^3 \times 144$ lattices at $10/g^2 = 6.72$. 
The strange and charm masses are set at their physical values while $m_l/m_s =
1/5$ for both lattices. The details of these configurations are summarized in
Ref.~\cite{HISQ_MILC}. We determined the lattice spacing by equating
the $\Omega(sss)$ baryon mass measured on these ensembles with its
physical value. The strange mass was tuned by setting the $\bar{s}s$
pseudoscalar mass to 685 MeV \cite{strange_tune}. 
%The effective mass for $\Omega(sss)$ baryon for the finer lattice are shown in Fig 1(b). 
The measured lattice
spacings are 0.0877(10) and 0.0582(5) fm for $32^3 \times 96$ and
$48^3 \times 144$ lattices respectively which are consistent with
0.0888(8) and 0.0582(4) fm as measured by MILC collaboration by
using the $r_1$ parameter~\cite{HISQ_MILC}.
The results reported here were obtained from 110
configurations on the coarser lattice, and 65 configurations on the
finer lattice. 
%Here we did not address the uncertainties in the determination 
%of lattice spacings which we will address in subsequent works.

For valence quarks we used the overlap action~\cite{overlap}. For the
numerical implementation of massive overlap fermions we followed the methods used
by the $\chi QCD$ collaboration~\cite{overlap_chiQCD}.
The low Wilson eigenmodes are projected out by the Arnoldi method
and the Zolotarev approximation is used to evaluate the sign function. 
We used the usual periodic
boundary condition in the spatial and antiperiodic in the temporal
directions. Gauge configurations were first fixed to Coulomb gauge 
and then smeared with a single level of HYP blocking. 
Using both point and wall sources we
calculated various point-point, wall-point as well as wall-wall
correlators.

Since for the charm quark $ma$ is not very small, we need to be careful
about discretization errors. The overlap action does not have $O(ma)$
errors. In order to estimate the size of discretization errors coming
from higher orders of $ma$, we look at the energy-momentum dispersion
relation of the 1S charmonia.  Expanding the energy momentum relation
in powers of ${\bf{p}}a$, one can write, for $|{\bf{p}}| << m_0, 1/a$
\begin{equation}
E(p)^2 = M_1^2 + {M_1\over{M_2}}{\bf{p}}^2 + O({\bf{p}}^4) = M_1^2
+{\bf{p}}^2c^2.
\end{equation} 
Here $M_1$ is the pole or rest mass $E(0)$, and $M_2$ is called
the kinetic mass ($M_1/c^2$). The difference between $M_1$ and $M_2$
is one measure of $\mathcal{O}(ma)$ cutoff effects. As highlighted in
Ref. \cite{ElKhadra:1996mp} (in the so-called Fermilab
interpretation), since $M_2$ controls the non-trivial physics of a
heavy hadron system, in using a relativistic action for heavy quarks,
one should use $M_2$ to measure the masses.

The charm mass is tuned by setting the spin-averaged 1S state mass,
$(m_{\eta_c}+ 3 m_{J/\psi})/4$, to its physical value, where we take
into account the kinetic mass, as defined above, in the definition of
mass. Previously~\cite{Basak:2012py} we tuned our charm mass with pole mass of mesons and
showed that the velocity of light ($c$) is not closer to 1, which leads to
$O(ma)^2$ errors.  We calculated pseudoscalar meson masses at various external
momenta $p^2 = (2 \pi/L)^2 n^2$, with $n \le 2$. We use wall sources at
finite momenta by putting a phase factor in the wall so as to
project to a particular momentum. This method is more suitable for
improving signal in correlators with finite
momenta~\cite{Basak:2012py}. In Fig.~1(a) we show $E(p)^2$ for various
momenta for the pseudoscalar meson on our finer lattices. The green
line is for the continuum dispersion relation, $E^2=m^2+p^2$, while
the blue line is the fitted dispersion relation with $c =
0.96(2)$. For coarser lattices we obtain $c = 0.92(3)$.
%=================
\begin{figure}[h]
\vspace{-0.1in}
%\begin{center}
\parbox{.45\linewidth}{
\centering
\includegraphics[width=0.5\textwidth,height=0.32\textwidth,clip=true]{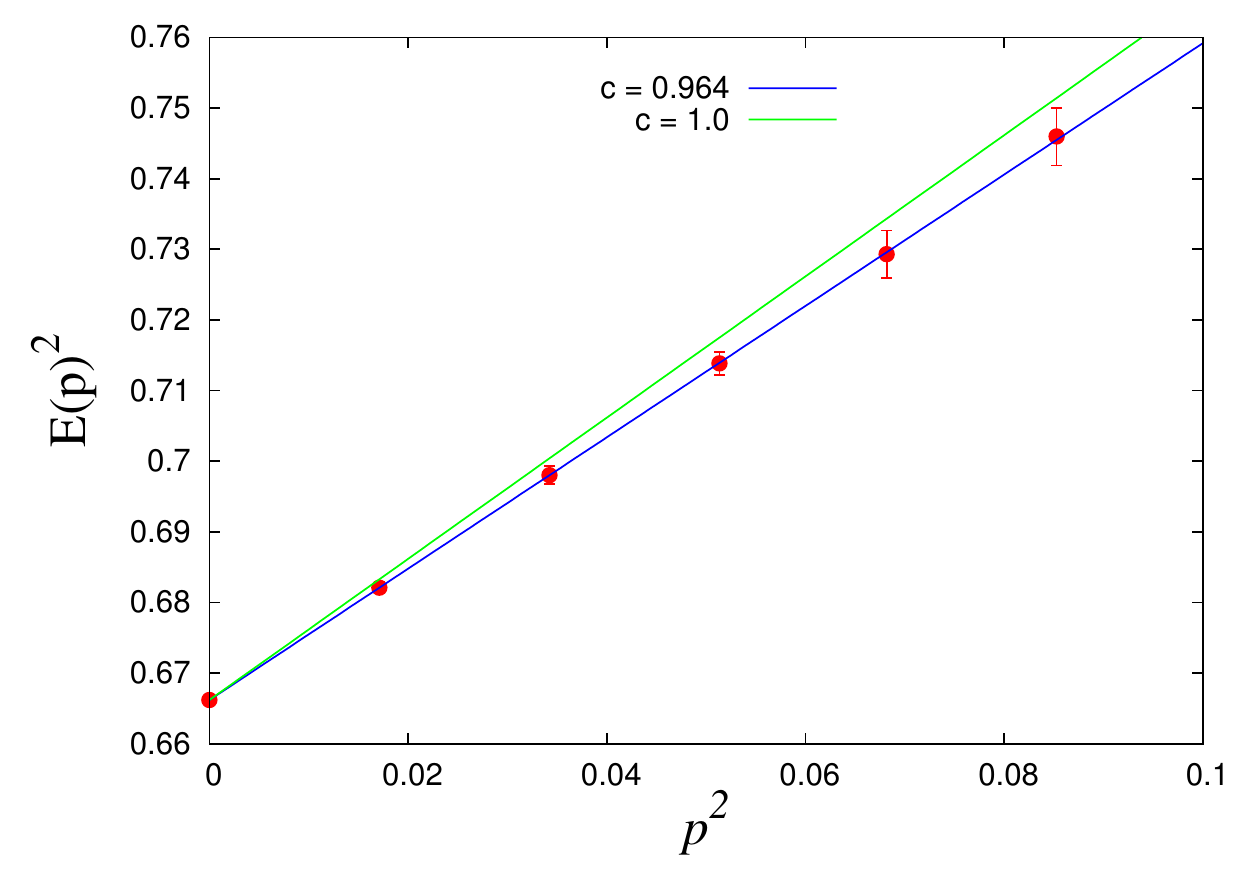}
(a)}
\hspace{0.75cm}
\parbox{.45\linewidth}{
\centering
\includegraphics[width=0.48\textwidth,height=0.31\textwidth,clip=true]{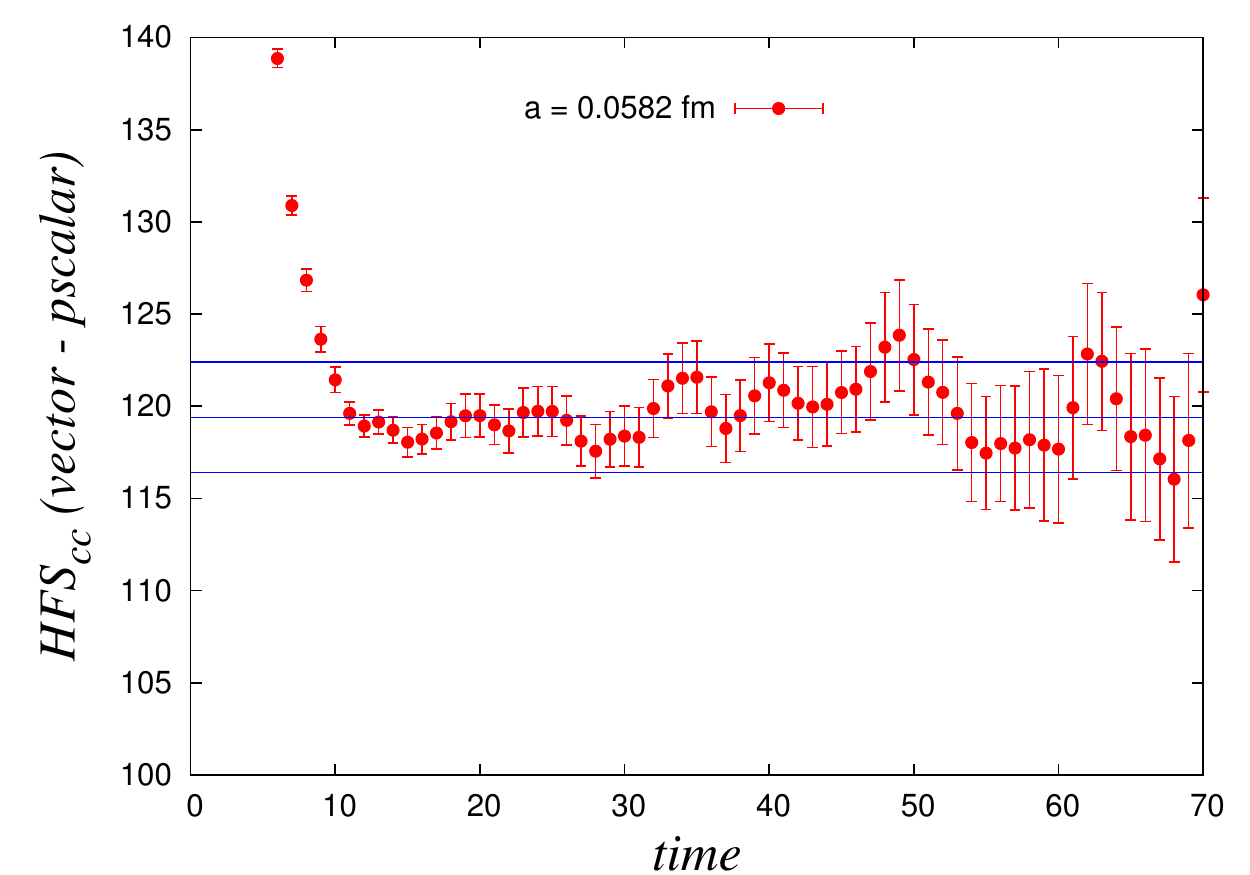}
(b)}
\vspace{-0.1in}
\caption{
(a) Energy-momentum dispersion relation for the pseudoscalar meson at charm mass on the finer lattices. Blue line is with c = 0.96(2) obtained by fitting our data while green line is with c = 1. (b) Effective hyperfine splitting in 1S charmonia for wall-point correlators for lattices with spacing 0.0582 fm. Horizontal lines show the fit results with one sigma error.}
\end{figure}

\vspace{-0.27in}
\section{Results}
\vspace{-0.08in}
Multimass methods help us to calculate the quark propagators over a wide
range of quark masses with $10-12\%$ overhead. Our extracted
pseudoscalar meson masses are within the range $400-5130$ MeV and
$230-4000$ MeV for the finer and coarser lattices respectively. 
In the following subsections we will discuss our results for mesons and baryons mainly in terms of energy splittings, as these have less systematic uncertainty as compared to extracted energies.

\vspace{-0.08in}
\subsection{Hyperfine splitting in 1S charmonia}
The hyperfine splitting in 1S charmonia is one of the most well
studied physical quantities in lattice charmonium calculations over
the years, and until very recently~\cite{MILC_hfs} lattice results were
found to be smaller than the experimental value ($\sim$ 116 MeV). This
underestimation is now understood to be mainly due to 
the discretization error associated
with the charm quark action and the quenched approximation.  In our
study we calculated this splitting. In Fig.~1(b), we show the
effective splittings between vector and pseudoscalar correlators
(jackknifed) at the tuned charm mass for wall-point correlators on
finer lattices. Horizontal lines shown are the fit results, with one
sigma errorbar.  Our final estimated results, for this hyperfine
splitting are $125(6)$ MeV and $119(3)$ MeV corresponding to coarser
and finer lattices respectively.
%%%%%%%%%%%%%%%%%%%%%%%%%%%%%%%%%%%%%%%%%%%%%%%%%%%%%%%%%%%%%%%
%\vspace*{-0.05in}
\subsection{Energy splittings in charmonia and charmed-strange mesons}
Beside 1S hyperfine splittings it is also important to consider energy
splittings between various other charmonia. In Fig.~2(a) we plot
energy splittings between axial, scalar and tensor charmonia from
pseudoscalar charmonium. In addition to this, we also calculated
charmed-strange mesons with various quantum numbers, and energy
splittings between these mesons are also plotted in Fig. 2(a). 
It is to be noted that tuning the charm mass by using kinetic mass has 
brought these splittings closer to experimental values than those previously obtained in Ref.~\cite{Basak:2012py}.
\begin{figure}[h]
\vspace{-0.1in}
%\begin{center}
\parbox{.45\linewidth}{
\centering
\includegraphics[width=0.5\textwidth,height=0.35\textwidth,clip=true]{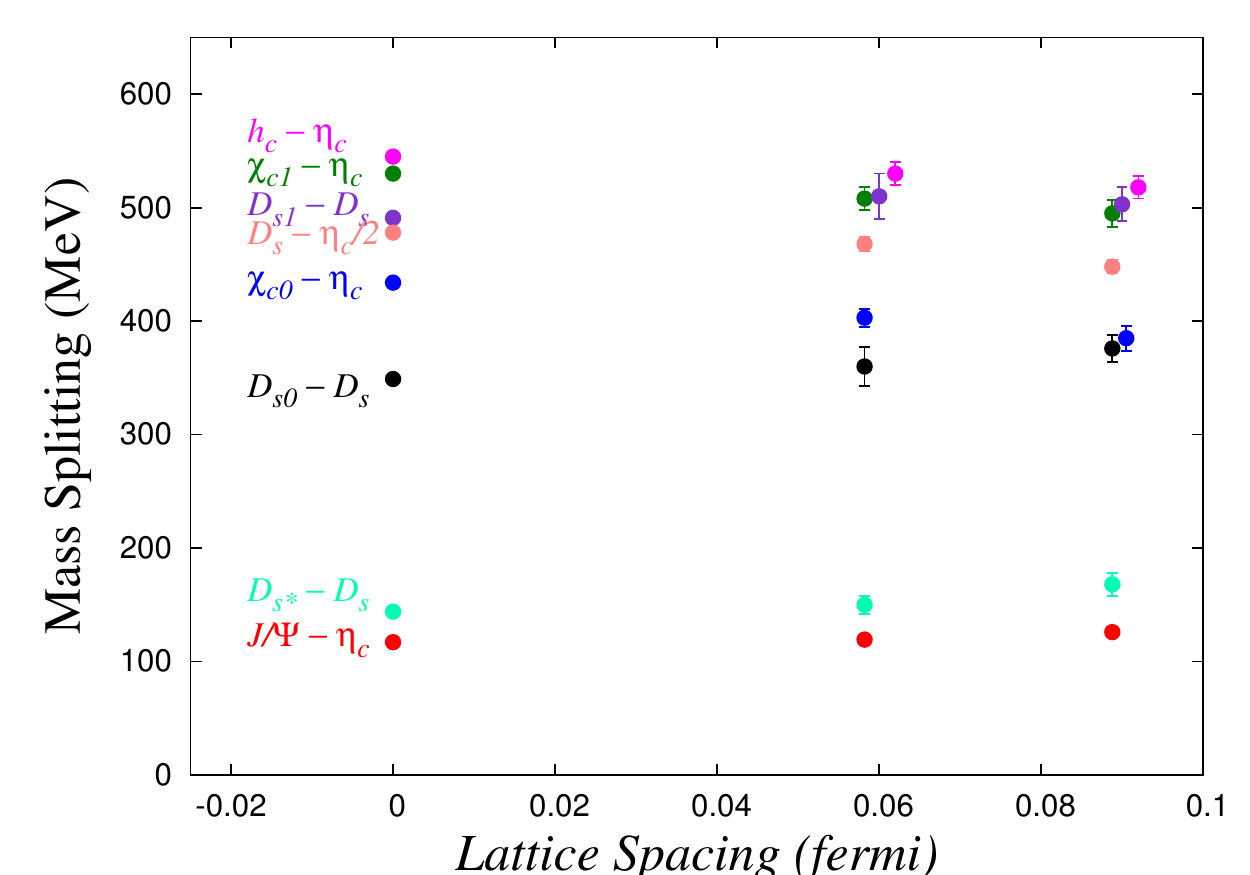}\\
(a)}
\hspace{0.75cm}
\parbox{.45\linewidth}{
 \centering
\includegraphics[width=0.45\textwidth,height=0.325\textwidth,clip=true]{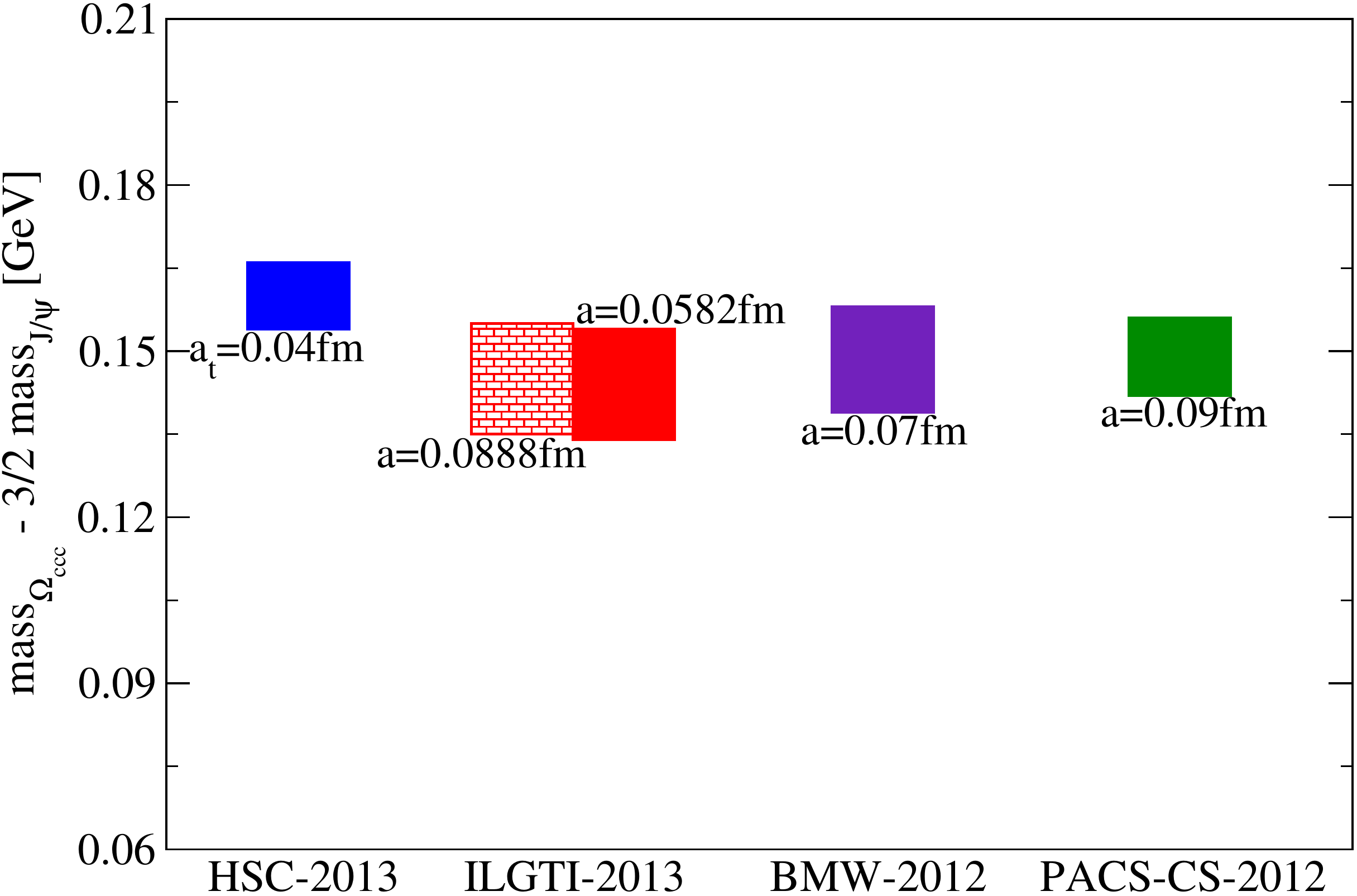}\\
(b)}
%\end{center}
\vspace{-0.1in}
\caption{(a) Meson mass splitting for charmonia and charmed-strange mesons at two lattice spacings. Experimental values are shown in the left side. (b) The mass splitting of $\Omega_{ccc} - {3\over2} J/\Psi$ along with other lattice results. Result from this work is shown in red colour.}
\end{figure}

\begin{figure}[h]
\vspace{-0.1in}
%\begin{center}
\includegraphics[width=0.5\textwidth,height=0.32\textwidth,clip=true]{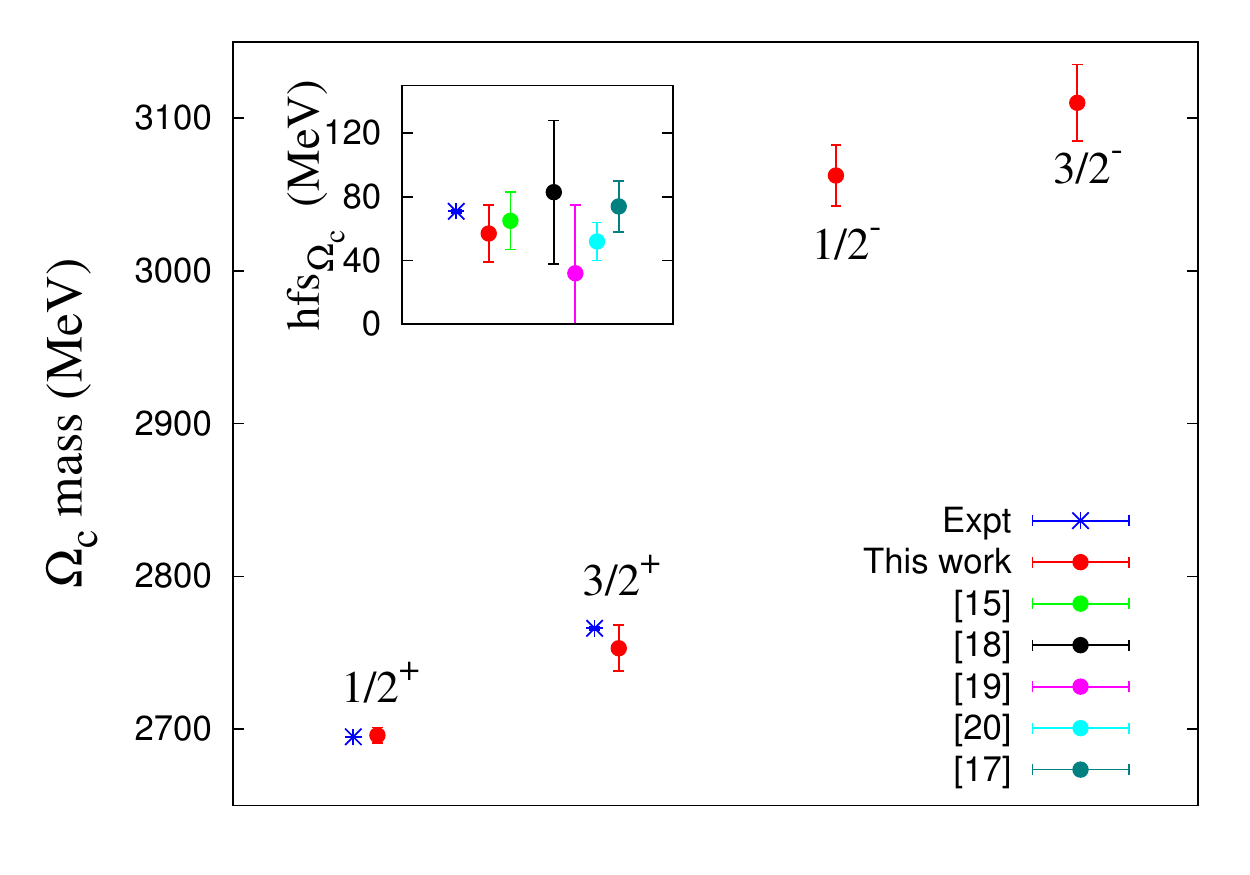}
\includegraphics[width=0.5\textwidth,height=0.32\textwidth,clip=true]{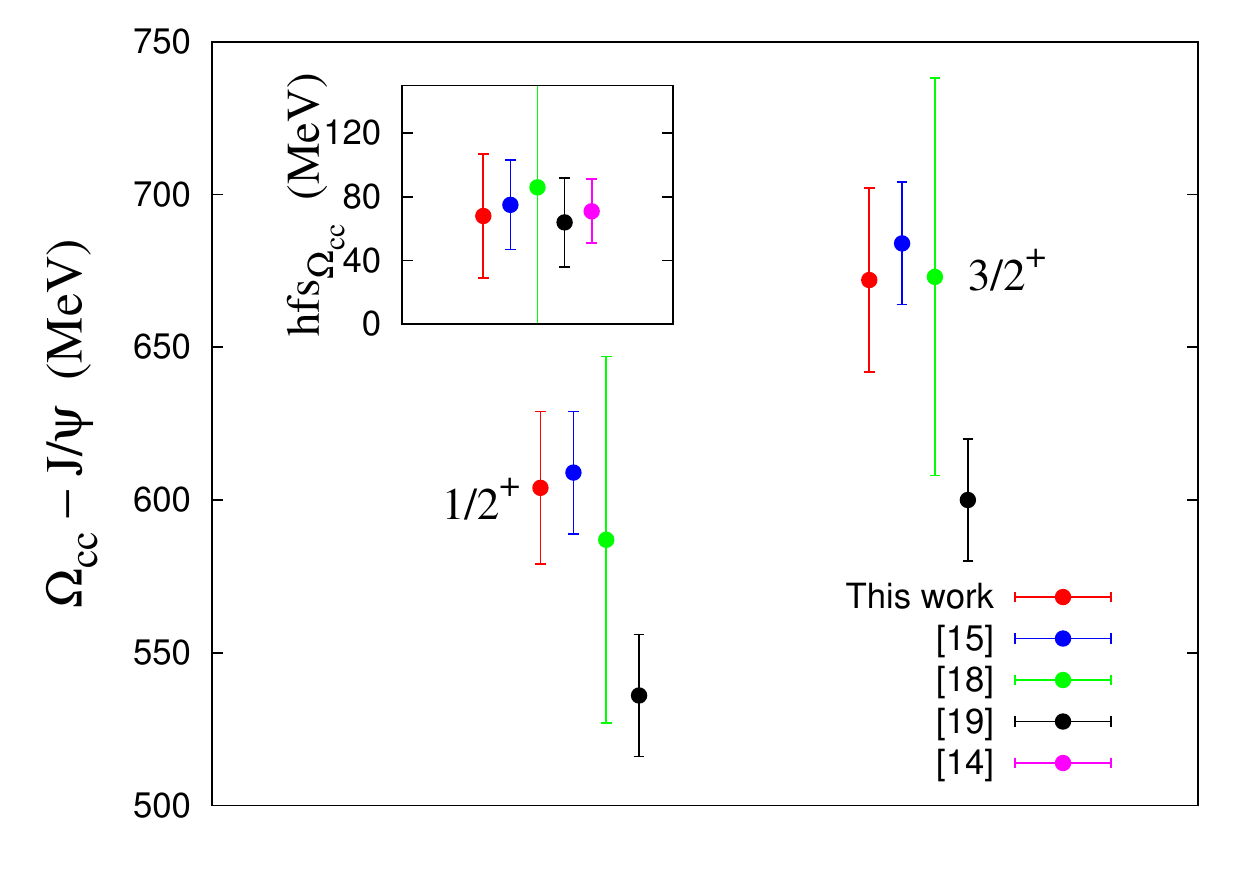}
%\end{center}
\vspace{-0.3in}
\caption{(a) $\Omega_c$ and (b) the mass splitting of $\Omega_{cc} -J/\Psi$. Inset figures are for hyperfine splittings between positive parity spin-3/2 and spin-1/2 states. Also shown are other lattice determinations, and the experimental values, where available.}
\end{figure}
%%%%%%%%%%%%%%%%%%%%%%%%%%%%%%%%%%%%%%%%%%%%%%%%%%%%%%%
\vspace{-0.2in}
\subsection{Charmed baryons}
Over the years the charmed mesons have been studied comprehensively 
and such studies have provided deeper understanding of 
the theory of strong interaction. 
However, the charmed baryons have not yet been studied in similar detail
though such studies can also provide similar inputs. 
%However, recently there have been much more focus to study heavy baryons. 
It is
thus crucial to study these baryons by using a first principle method
 such as lattice QCD. In this work we extracted ground state spectra of
charmed baryons with one or more charm quark content, for example,
baryons with quark content $csu, cuu, css$, $ccs$, and $ccc$. The
study of a particular baryon which draws immediate attention is the
triply-charmed $\Omega(ccc)$, a baryonic analogues of charmonia, which
according to Bjorken~\cite{Bjorken:1985ei}, may provide a new window
for understanding the structure of baryons.  Though the theory of
strong interaction unambiguously predicts such a state, similar to its light
quark counterpart $\Delta(uuu)$ and $\Omega(sss)$, it has not yet been
observed. In Fig.~2(b) we plot the mass splitting of
$\Omega_{ccc} - {3\over2} J/\Psi$. A factor 3/2 is included to account
for the difference in the charm quark content 
in $\Omega_{ccc}$ and $J/\Psi$, and thus
this splitting mimics the binding energy for such a state. We also
plotted other lattice determinations~\cite{Padmanath:2013zfa,Namekawa:2013vu,Durr:2012dw} for this quantity and our result
is consistent with those. In Fig.~3 we showed results for
$\Omega_c(css)$ and $\Omega_{cc}(ccs)$ baryons, and for the later case
energy splittings of $\Omega_{cc} -J/\Psi$.  
It is to be noted that
for these baryons we extracted masses for both spin 1/2 and spin 3/2
with both parities, some of which are yet to be measured
experimentally.
In the inset we also plot the 
hyperfine splittings between positive parity spin-3/2 and spin-1/2 states.
 Our results are consistent with other lattice
results~\cite{Namekawa:2013vu,Mathur:2002ce,Briceno:2012wt,Bali:2012ua,Padmanath:2013bla}, and the experimental values, where
available.  We are in the process of adding another lattice spacing and after that we will we will carry out both continuum as well as
chiral extrapolations by using mixed action
partially quenched chiral perturbation theory~\cite{MAPQchpt}. One
also needs to evaluate $\Delta_{mix}$ \cite{delta_mix}, the low energy
constant representing $\mathcal{O}(a^2)$ discretization dependence.

%%%%%%%%%%%%%%%%%%%%%%%%%%%%%%%%%%%%%%%%%%%%%%%%%%%%%%%%
%%%%%%%%%%%%%%%%%%%%%%%%%%%%%%%%%%%%%%%%%%%%%%%%%%%%%%%%
\vspace*{-0.1in}
\subsection{Decay constants}
\begin{figure}[h]
\vspace{-0.1in}
%\begin{center}
\includegraphics[width=0.52\textwidth,height=0.32\textwidth,clip=true]{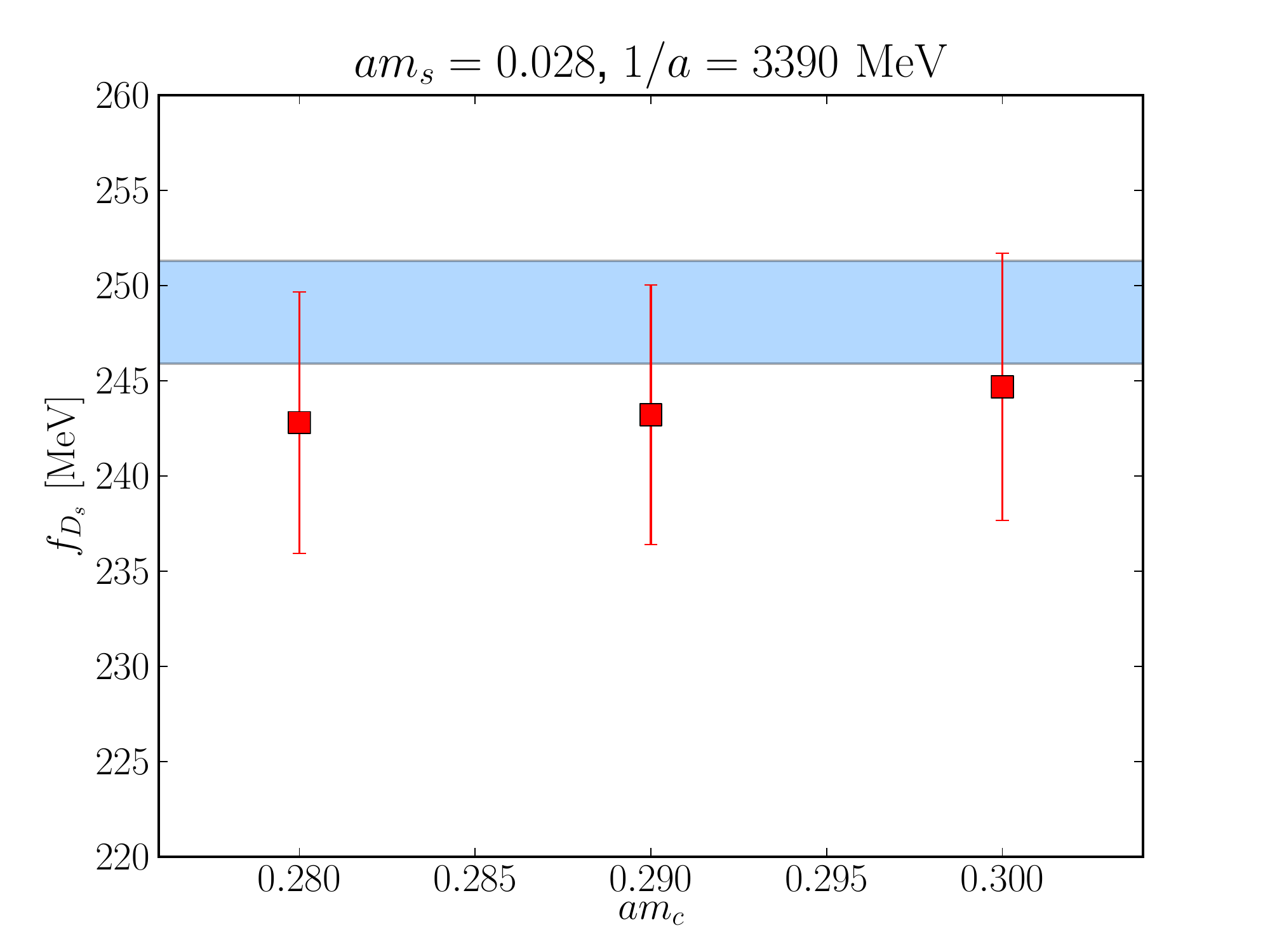}
\includegraphics[width=0.52\textwidth,height=0.32\textwidth,clip=true]{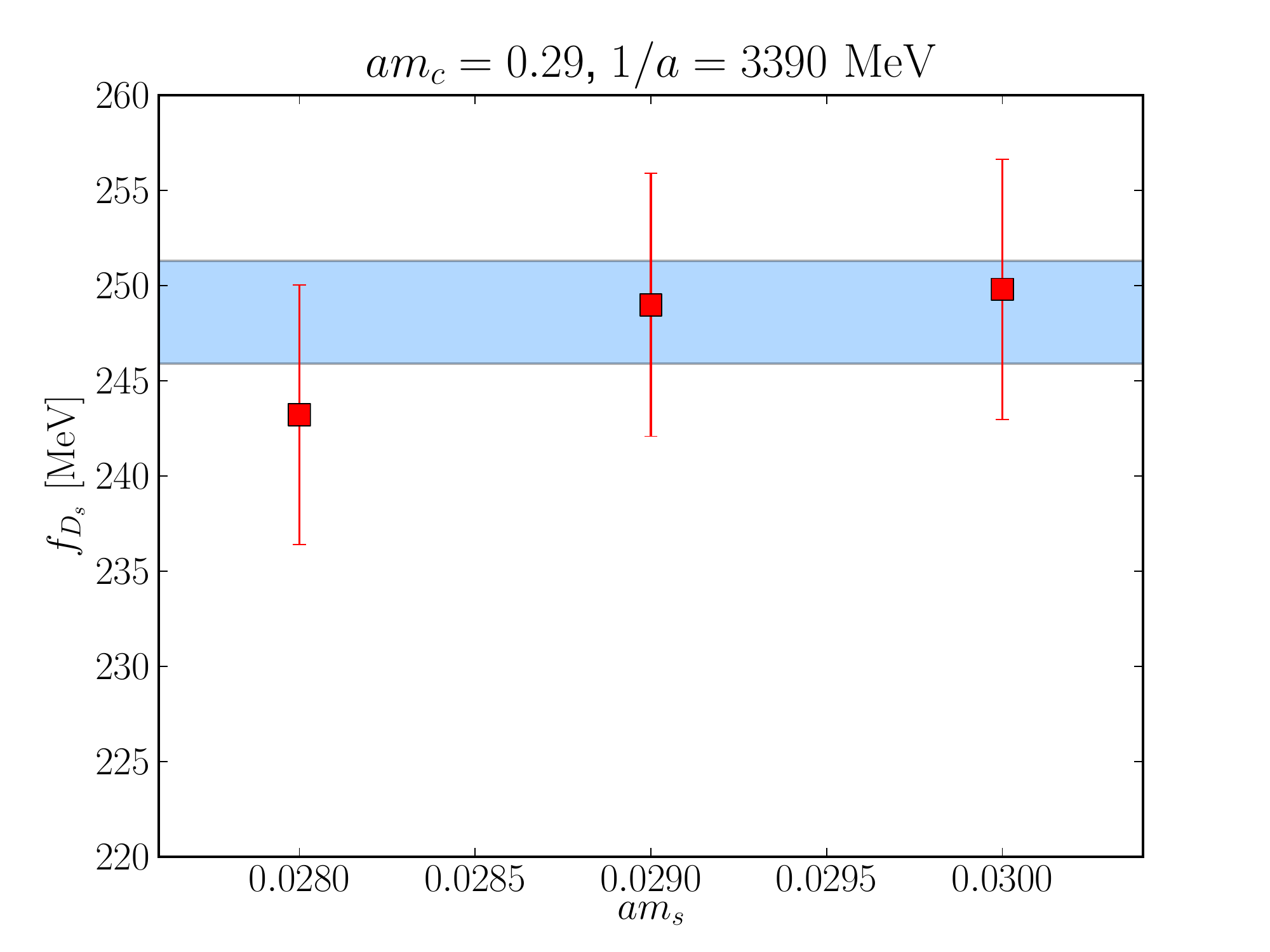}\\ 
\includegraphics[width=0.52\textwidth,height=0.32\textwidth,clip=true]{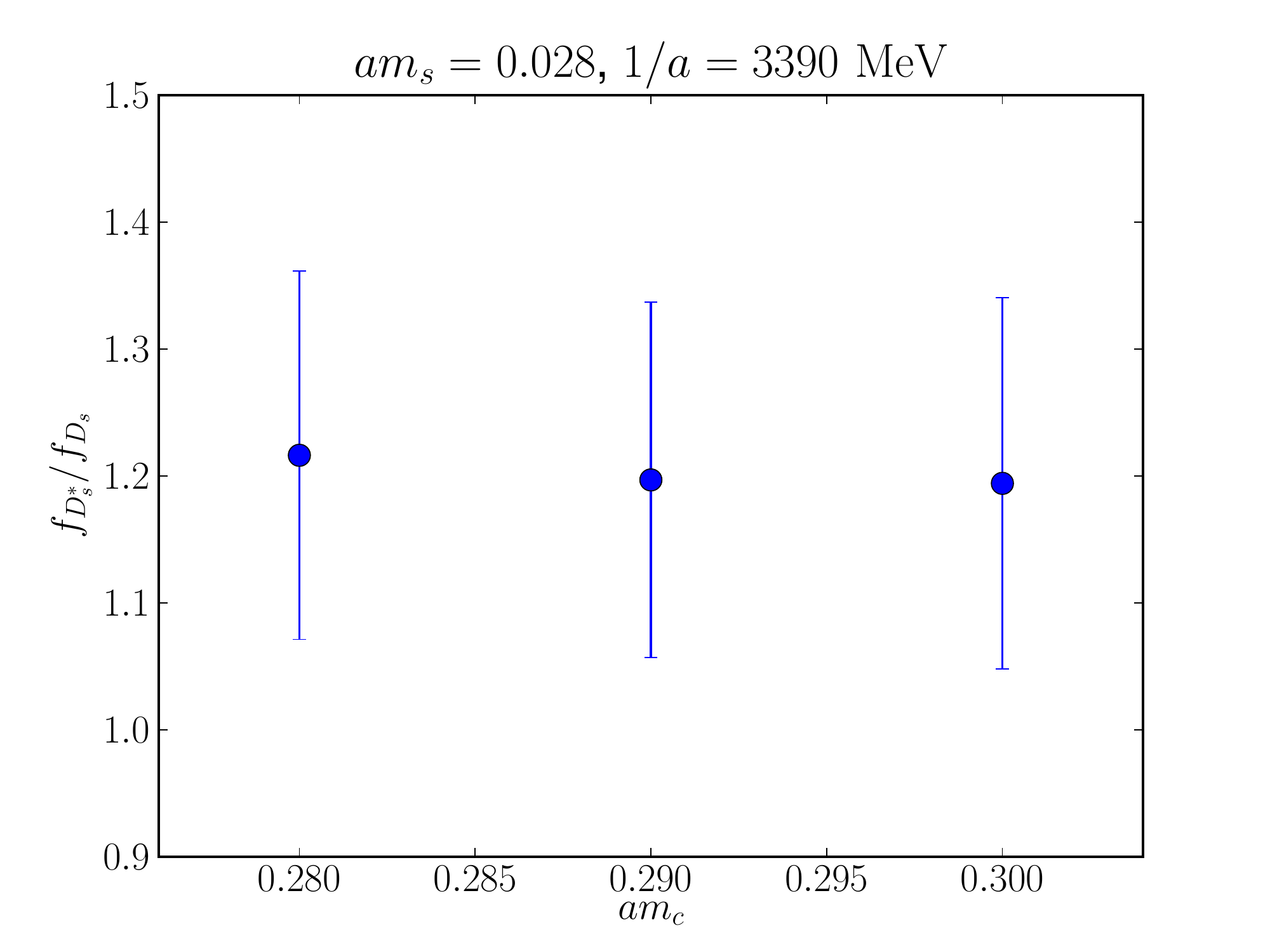}
\includegraphics[width=0.52\textwidth,height=0.32\textwidth,clip=true]{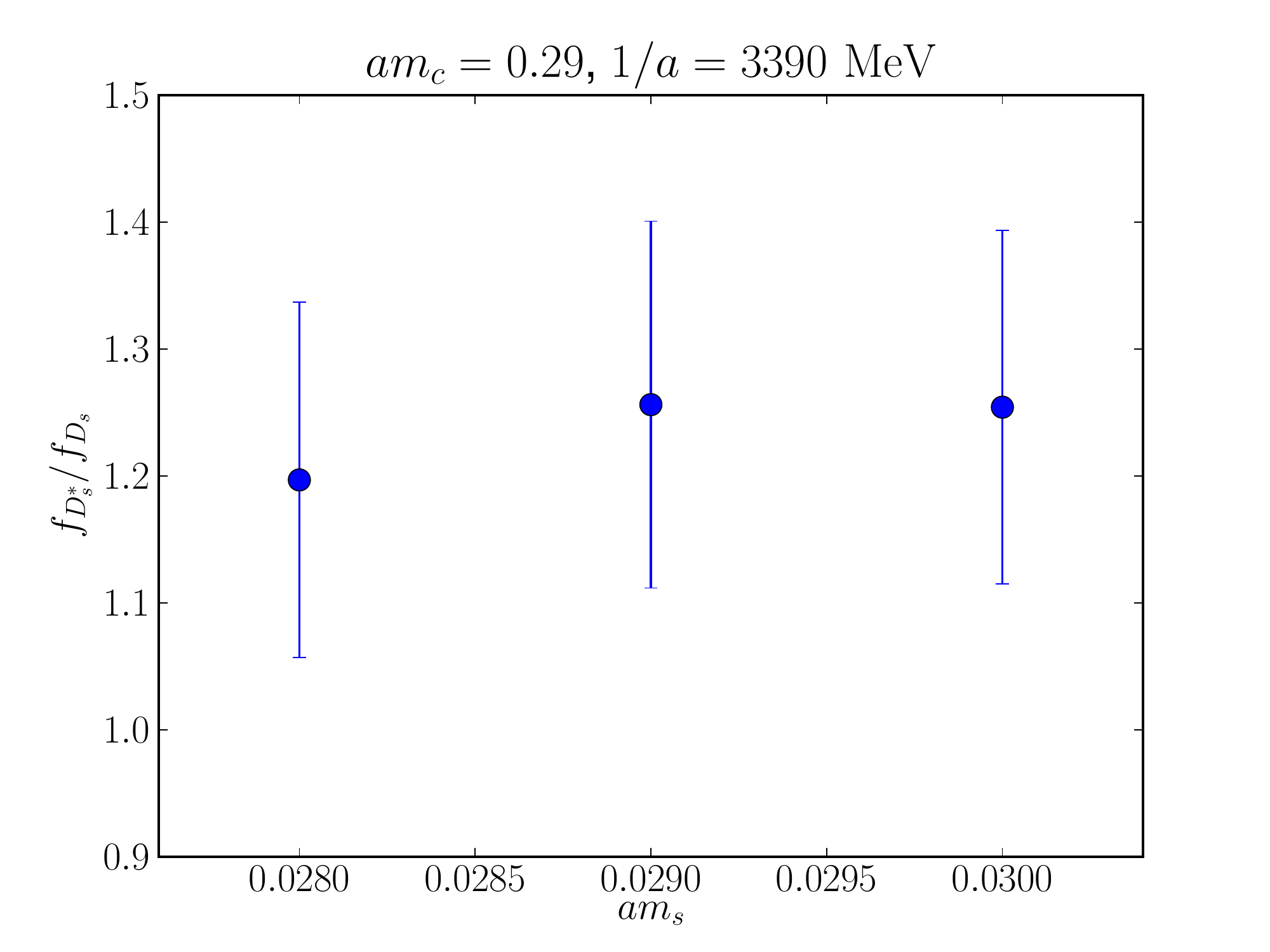}
%\end{center}
\vspace{-0.3in}
\caption{(Top) $f_{D_s}$ as a function of charm and strange input masses on the 
finer of our two ensembles ($a^{-1} \approx 3.4 \mbox{ GeV}$).  The physical result quoted
in the PDG is shown as a blue band.
(Bottom)  Results for the ratio $f_{D_s^*}/f_{D_s}$, assuming $Z_A = Z_V$.}
\label{fDs_ratio}
\end{figure}
The decay constants of heavy-light mesons are experimentally very important:
they are essential ingredients in extracting CKM matrix elements from decays
of heavy-light mesons.
Here we present preliminary results for the pseudoscalar decay constant $f_{D_s}$,
as well as the ratio of vector to pseudoscalar decay constants $f_{D_s^*}/f_{D_s}$.
%These are important quantities, for example in determining the CKM matrix element $|V_{cs}|$ from experimental decay rates.
The decay constants $f_{D_s}$ and $f_{D_s^*}$ are defined by
\begin{equation}
\langle 0| A_\mu |{D_s(p)} \rangle = f_{D_s} \,p^{\mu}, \qquad 
\langle 0| V_\mu |{D_s^*(p, \lambda)}\rangle  = f_{D_s^*} \, M_{D_s^*} \, \epsilon^{\lambda}_\mu
\label{decay_const_def}
\end{equation}
where $\epsilon^{\lambda}_\mu
$ is a polarization vector, and $A_\mu$ and $V_\mu$ are the continuum currents related to lattice operators
by $\{A_\mu, V_\mu\} = \{Z_A \bar{s} \gamma_\mu \gamma_5 c, Z_V \bar{s} \gamma_\mu c \}$.
The $D_s$ decay constant is determined from the relation
\begin{equation}
M^{2}_{D_s} \, f_{D_s} = (m_c + m_s) \langle 0| P |{D_s}\rangle \,,
\label{PCAC_rel}
\end{equation}
where $P=\bar{s} \gamma_5 c$.  Note that the quantities on the right are bare quantities.

We are currently in the process of computing $Z_V$ using sequential propagators, which 
will allow us to determine $f_{D_s^*}$.
Since we expect that $Z_A/ Z_V \approx 1$
for our chiral action (for massless fermions $Z_A/Z_V = 1$),
here we present results
for the quantity $\frac{Z_A f_{D_s^*}}{Z_V f_{D_s}} \approx \frac{f_{D_s^*}}{f_{D_s}}$.  

The matrix elements in Eqs.(3.1)
and (3.2) are determined in the standard way by fitting two-point
correlation functions $\langle O^\dagger(t) O(0)\rangle$ where $O =
\{A_4, V_4, P\}$. The results presented here are all obtained using Coulomb gauge-fixed
point-source propagators.
We are also exploring point-wall and wall-wall correlators as well as different combinations
of operator (e.g.\ $A_4$-$P$).
All uncertainties are computed via single-elimination jackknife.

In Fig.~\ref{fDs_ratio}, we show results for $f_{D_s}$ on the finer of our two ensembles ($a^{-1} \approx 3.4 \hbox{GeV}$) as both the heavy mass $m_c$ and light mass $m_s$ are varied.
We see little variation in the results over the ranges studied, which are consistent with the 
value given in the PDG.  Fig.~\ref{fDs_ratio} also shows results for $f_{D_s^*}/f_{D_s}$,
where we have assumed that $Z_A = Z_V$. It is expected that the mixed
action effects will be smaller for heavy-light mesons, and in the
ratio its effects will be minimal.

%%%%%%%%%%%%%%%%%%%%%%%%%%%%%%%%%%%%%%%%%%%%%%%%%%%%%%%%%%%%%%%%
\vspace{-0.12in}
\section{Conclusions}
\vspace{-0.06in}
In this work we reported preliminary results on the ground state
charmed hadron masses along with charmed-strange meson decay constants by using
a mixed action approach, comprising overlap valence quarks, generated
on the background of dynamical 2+1+1 flavours HISQ configurations.
The results, in particular the hyperfine splitting of 1S charmonia,
are encouraging and suggest that the overlap valence on 2+1+1
flavor HISQ configurations is a promising approach to do lattice QCD
simulation with light, strange and charm quark together in the same
lattice formulation. Discretization errors of the overlap action for the charm quark are reduced by tuning the charm quark mass with kinetic mass, rather than pole mass, as suggested in the Fermilab formulation of heavy quarks~\cite{ElKhadra:1996mp}.

This is a continuing study and we expect to be able to do suitable
chiral and continuum extrapolations, to make experimentally relevant
predictions for various charmed
baryons. The splitting ($m_{\Omega_{ccc}}-{3\over2}m_{J\//\Psi}$),
between $J\//\Psi$ and the unknown triply-charmed baryon
$\Omega_{ccc}$ was found to be $145(10)$ MeV and $144(10)$
MeV, on our coarser and finer lattices respectively. 
We are also studying heavy-light decay
constants and in the process of calculating renormalization constants.

%%%%%%%%%%%%%%%%%%%%%%%%%%%%%%%%%%%%%%%%%%%%%%%%%%%%%%%%%%%%%%%%
 \vspace{-0.12in}
 \section{Acknowledgement}
The computations were carried on the Blue Gene P of Indian Lattice Gauge Theory
Initiative, Tata Institute of Fundamental Research (TIFR), 
Mumbai. We would like to thank A. Salve and K. Ghadiali for technical support. We are
grateful to the MILC collaboration and in particular to S. Gottlieb, for
providing us with the HISQ lattices.

\end{document}